# Dante, astrología y astronomía

Alejandro Gangui

Instituto de Astronomía y Física del Espacio, CONICET,
Centro de Formación e Investigación en Enseñanza de las Ciencias, FCEyN-UBA

Hijo de Alighiero di Bellincione y de Bella (o quizás Gabriella), Dante Alighieri (o quizás "Durante" Alighieri, en su forma completa original) nació en 1265 en Florencia, capital de la Toscana. La fecha exacta no ha llegado a nosotros, ya que lo poco que sabemos sobre su vida es lo que quedó escrito en diversos documentos de la época, en comentarios de otros escritores y en sus propias obras literarias. En estas últimas, Dante no se mostró demasiado preciso.

Sin embargo, en el canto XXII del Paraíso, el tercer cántico de su obra cumbre, la *Divina Comedia*, Dante declaró haber nacido bajo la influencia de Géminis, conforme a su signo astrológico, y siguiendo una tradición antiquísima aunque de dudosa reputación actual. Dante, en esta parte de su obra, cantó a las estrellas que forman la constelación de los Gemelos, las que describió así (versos 112-117; traducción de Ángel J. Battistessa):

*"O gloriose stelle, o lume pregno*
 *di gran virtú, dal quale io riconosco*
 *tutto, qual che si sia, il mio ingegno,*

*con voi nasceva e s'ascondeva vosco*
 *quegli ch'è padre d'ogni mortal vita,*
 *quand'io senti' di prima l'aere tosco;"*

*"¡Oh gloriosas estrellas, lumbre henchida*
 *de gran virtud, en la que reconozco*
 *todo, sea cual fuere, el propio ingenio!*

*Con vosotras nacía y se ocultaba*
 *el que es padre de toda mortal vida,*
 *cuando niño sentí el aire toscano;"*

Debemos admitir que poca gente se escribió un acta de nacimiento más poética y original… Lo que nos quiere trasmitir Dante con estos versos es que en conjunción con la constelación de los Gemelos surgía y se ocultaba aquel astro que genera todas las cosas terrenas, el Sol, cuando el poeta respiró por vez primera –en el momento de su nacimiento– el aire de la Toscana. De este y otros datos (ver también el canto XV del Infierno, versos 55-57) podemos deducir entonces que Dante debió haber nacido entre el 22 de mayo y el 21 de junio de 1265. La fecha es aún tema de discusión entre los estudiosos: Battistessa, por ejemplo, la ubica entre el 14 de mayo y el 15 de junio.

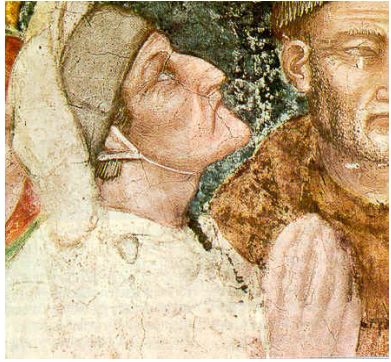

**Retrato de Dante en el fresco de Nardo di Cione "El juicio universal",
Capilla Strozzi, Basílica de Santa María Novella, Florencia.**

La lectura de versos en que el poeta hace referencia a la posición de los astros en el día de su nacimiento lleva a preguntarse si creía en la astrología. Curiosamente, la respuesta es tanto afirmativa como negativa.

En la Edad Media, astronomía y astrología se diferenciaban mal. En el siglo XIII eran raros los filósofos de la naturaleza que no tenían genuino interés por supuestas relaciones "causales" entre el cielo y las vicisitudes terrestres. Sin embargo, para Dante no existía ambigüedad, pues, como lo afirma en el segundo tratado de su gran trabajo filosófico *El Banquete* (*Convivio*, capítulo XIII) :

*"*[La astronomía] *es la más alta de entre toda*s [las ciencias] *ya que, como lo afirma Aristóteles en el comienzo del Alma, la ciencia es alta en nobleza por la nobleza de su sujeto y por su certeza; y ésta, más que cualquiera de las otras, es noble y alta, por el noble y alto sujeto de su estudio, ya que trata sobre el movimiento del cielo; y alta y noble por su certeza, que no tiene defecto alguno, como procedente de un muy perfecto y regulado principio. Y si algunos creen ver defecto en ella, el defecto no está de su parte, sino que se debe, como dice Ptolomeo, a nuestra negligencia, y es a esta última que debe ser imputado."*

En este extracto, Dante hace mención del célebre texto *De anima* (*Sobre el Alma*) en el que Aristóteles tratara de explicar qué es el alma o "el principio de la vida" que poseen tanto plantas, como animales y seres humanos.

Para Dante, la astrología, de acuerdo con las directivas de San Agustín y de los demás Padres de la Iglesia, debía ser condenada; los astrólogos contemporáneos de Dante aparecerán en el primer reino de la *Commedia*: el Infierno.

Para nuestro poeta, la principal motivación de esta condena era la necesidad de salvaguardar el libre albedrío humano: si fuese verdad que existe una relación entre el alma del hombre y los astros, esto arruinaría su libertad. ¿Para qué actuar bien y ser un buen cristiano, si todo estaba ya "escrito" en los astros? Dante, un hombre culto de su tiempo pero embebido en una religiosidad profunda, rechaza esta "conexión astral". Sin embargo, veremos que –no ya para el alma, pero sí para las cosas terrenas– Dante admite con ciertas sutilezas el poder de los astros (ver recuadro *Una hipótesis "fantástica"*).

Pasemos ahora a los signos. Todos sabemos cuál es nuestro signo de nacimiento, aunque por supuesto algunos lo tienen más presente que otros. Dante, como vimos, se declara Géminis, esto es, afirma en varias ocasiones haber nacido bajo el signo de los Gemelos.

Si preguntamos a alguien (a quien no le desagraden los horóscopos) cuál es su signo, muy probablemente nos responda de inmediato. Podemos entonces sugerirle que nos diga *su verdadero signo de nacimiento*, y explicarle que los signos habituales que se publican en periódicos y revistas generalmente *no* son los correctos, a menos que uno haya nacido un par de miles de años atrás. La intriga de nuestro hipotético interlocutor permitiría explicar algunas cuestiones interesantes de astronomía, por ejemplo, las consecuencias que tiene para la astrología la existencia de la precesión de los equinoccios (ver recuadro *El movimiento de los cielos*).

Muchos lectores de horóscopos no tienen presente que el signo astrológico de una persona indica que en el momento de su nacimiento el Sol, visto desde la Tierra, estaba (o "residía") en el cielo entre las estrellas de la constelación definida por ese signo. Si preguntamos entonces cuál es el mejor momento para observar en el cielo la constelación de ese signo, la respuesta usual será el mes en que la persona cumple años. Ese, sin embargo, es el peor momento. En ese mes, justamente, el Sol y las estrellas de la constelación se hallan en la misma parte del cielo (mirando desde la Tierra, por supuesto). Dicha constelación aparece de día, las estrellas que la forman están detrás del Sol y por ello *no* pueden verse... Y esto ya lo mencionó Dante unas páginas atrás para "sus estrellas": *"Con vosotras nacía y se ocultaba / el que es padre de toda mortal vida, / cuando niño sentí el aire toscano"*. Así, el mejor momento para mirar nuestra constelación zodiacal de nacimiento en el cielo nocturno es unos seis meses antes (o después) del cumpleaños.

A lo largo del año, por la traslación de la Tierra alrededor del Sol, visto desde nuestro planeta este se va desplazando entre las bien conocidas doce constelaciones zodiacales. El camino o trayectoria anual del Sol en el cielo (distinto del camino diario del Sol en el firmamento, producto de la rotación de la Tierra sobre su eje) se denomina la *eclíptica*. ¿Correcto hasta aquí? No exactamente.

Ante todo, no hay doce constelaciones a lo largo del camino del Sol sino trece. La constelación adicional ubicada sobre la eclíptica se llama *Ofiuco*, personaje mitológico que en la antigüedad representaba el "serpentario" o "portador de serpientes" (literalmente *Ofi-Okos*, en Griego). Y *la trayectoria* que sigue el Sol se ubica sobre Ofiuco entre el 30 de noviembre y el 17 de diciembre (véase la *tabla adjunta*). En consecuencia, el signo zodiacal de quien haya nacido entre estas fechas es Ofiuco y no Sagitario.

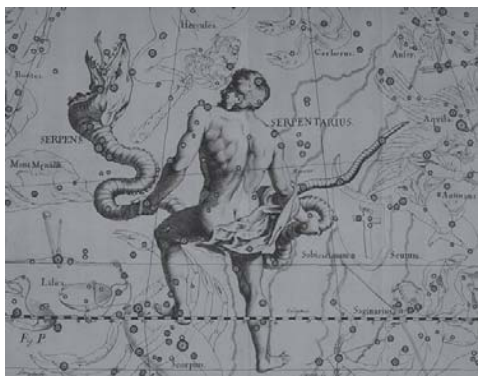

**La constelación de Ofiuco o el portador de serpientes, en la *Uranographia totum coelum stellatum* de Johannes Hevelius, 1690. La línea imaginaria que pasa por el talón izquierdo del personaje mitológico corresponde a la eclíptica, el "camino" del Sol sobre el fondo de las estrellas del cielo. Nótese que aquí la representación de las constelaciones está hecha como si mirásemos a la esfera celeste "desde afuera" y *no* desde su interior como realmente sucede en la práctica (en aquella época, los globos celestes comenzaban a aparecer en las casas reales y nobles en toda Europa) y por ello la orientación del serpentario (y de todas las demás figuras de la *Uranographia* de Hevelius) está invertida con respecto a la que podemos ver en el cielo nocturno.**

Apuntemos también que a veces se menciona a Cetus, la Ballena (adviértase la misma raíz latina en cetáceo), como nueva integrante de las constelaciones del zodíaco (lo que elevaría el número a 14 –y no solo a 13– constelaciones zodiacales). En este caso hay que hilar un poco más fino. Pues los límites actuales impuestos por los astrónomos a las constelaciones del cielo muestran que una parte de Cetus se acerca mucho a la eclíptica, pero que no llega a tocarla. Sin embargo, el Sol no es un punto en el cielo, sino un disco cuyo diámetro aparente es de unos 30 minutos de arco. La medida parece insignificante (menos de 0,3% de una semicircunferencia trazada sobre la bóveda celeste), hasta que se advierte que la separación angular entre el límite de Cetus y la trayectoria del Sol es de unos nueve minutos de arco.

El resultado es que el 27 de marzo, día del año de mayor acercamiento de la eclíptica a Cetus, durante algunas horas (unas 12 horas aproximadamente) una parte pequeña del Sol escapa temporariamente los confines de Piscis (los Peces) e incursiona en la constelación de la Ballena. ¿Alcanza esto para incluirla en la tabla? A juzgar por el número de personas que anualmente nacen durante esas horas, la respuesta debería ser afirmativa. Este autor, sin embargo, sugiere dejar a Cetus fuera de la lista. Entre otras razones, porque el número de días en cuestión sería fraccionario; además habría que anotar a Piscis dos veces, pues después de pasearse por Cetus esa fracción del disco solar vuelve a entrar en la constelación de los Peces.

Luego, no incluimos a Cetus en nuestra tabla adjunta. Pero tengamos en cuenta que la astrología no solo se interesa por el Sol sino también por las posiciones de los planetas, y éstos describen trayectorias aparentes en el cielo de la Tierra que se apartan un poco del camino solar. Aunque el Sol en la actualidad no permanece por mucho tiempo en el "área de influencia" de Cetus, muchos de los planetas sí lo hacen (por ejemplo, Mercurio, entre el 6 y el 7 de abril de 2008, o Venus, del 11 al 13 de abril de 2008) y por eso algunos astrólogos toman en cuenta catorce constelaciones en sus "cálculos".

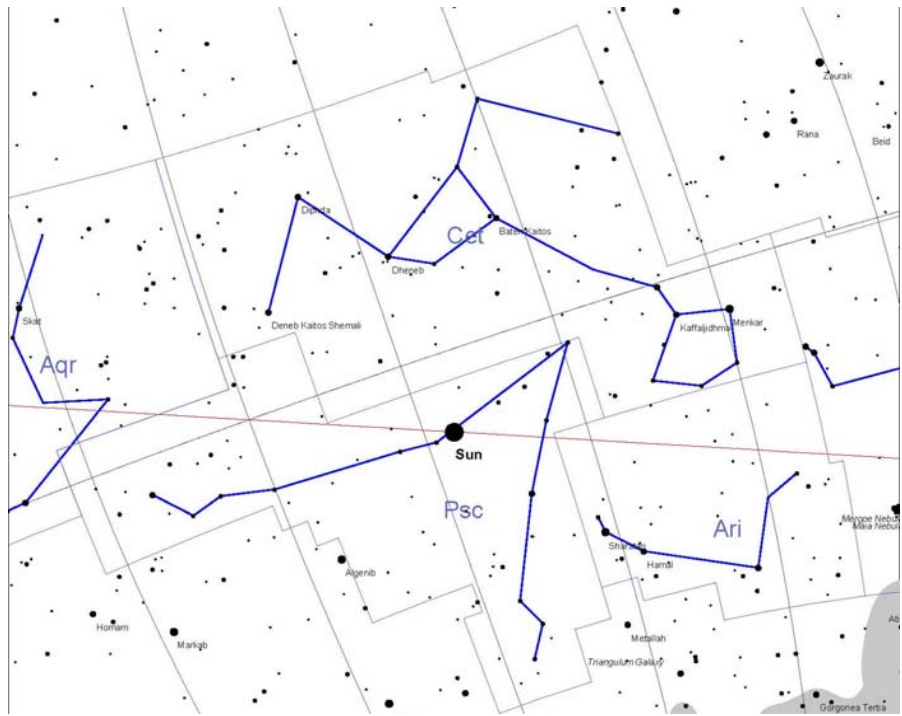

**Trayectoria aparente del Sol (la "eclíptica", en rojo) proyectada sobre la bóveda celeste de la Tierra, con la ubicación del Sol correspondiente al día 8 de abril de 2008 (al mediodía de Buenos Aires). En este momento el astro se halla en la constelación de Piscis (Psc) y se encamina hacia Aries. Un mes antes el Sol se encontraba en dominios de Acuario (Aqr). Las líneas azules más débiles indican las divisiones acordadas por los astrónomos para las diferentes constelaciones del cielo. Nótese en esta imagen la región hacia la izquierda del Sol, donde la eclíptica parece tocar el borde de la constelación de la Ballena o Cetus (Cet). En esa ubicación se halla el Sol el día 27 de marzo. Una imagen ampliada muestra, por el contrario, que ambas líneas no se tocan (ver figuras siguientes). Sin embargo, el radio aparente del disco solar es mayor que la distancia entre las líneas roja y azul débil. En consecuencia, una fracción del Sol ingresa en Cetus durante algunas horas. En esta imagen, obtenida para la ciudad de Buenos Aires con un programa de cálculo astronómico, el Sur se encuentra hacia arriba y a la izquierda.**

A la luz de lo dicho sobre la posición del Sol en los distintos meses del año, y de que, en realidad, no son doce sino por lo menos trece las constelaciones zodiacales, ¿por qué el signo que anuncian periódicos y revistas no coincide con el verdadero? Porque con el andar de los siglos las constelaciones del Zodíaco se han desplazado. Ello se debe a un lento movimiento del eje de rotación de la Tierra que causa la *precesión de los equinoccios*, según explicamos en el recuadro *El movimiento de los cielos*.

Este desplazamiento de las constelaciones del Zodíaco no es tomado en cuenta por la gran mayoría de los astrólogos populares (cultores de la llamada *astrología tropical*), pues afirman que hace varios miles de años hubo coincidencia entre signos y constelaciones, y que, por ello, el área de la bóveda celeste abarcada por cada constelación en ese entonces debió quedar con la "influencia" (o con el "signo") de los respectivos astros. No les interesan las constelaciones sino, más bien, las posiciones que ellas tenían en la época de los babilonios. Sin embargo, un número reducido de astrólogos prefirió *aggiornarse* y tomar en cuenta la precesión y, con ésta, también el lento pero inexorable movimiento de las constelaciones. Ellos practican la denominada *astrología sidérea*, igual de esotérica que la primera, pero con la aspiración de tomar en cuenta algunas conclusiones científicas.

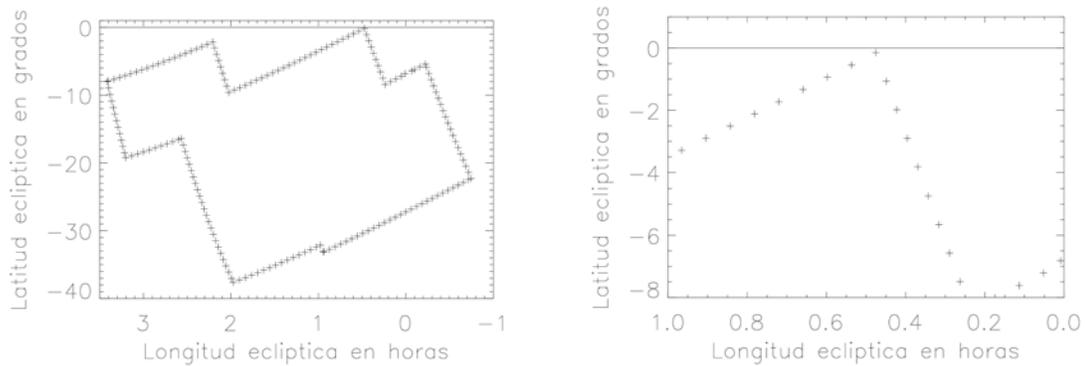

**El panel izquierdo muestra los límites de la constelación de la Ballena a partir de los datos de *Constellation Boundary Data* (disponibles en ftp://cdsarc.u-strasbg.fr/cats/VI/49/), donde se transformaron las coordenadas ecuatoriales (ascensión recta y declinación) a latitud y longitud eclípticas, válidos para el equinoccio 2000. Se usó un valor de 23,4392911º para la oblicuidad de la eclíptica respecto al ecuador celeste, como es recomendado por la IAU. La figura es similar a la anterior (sin las estrellas que forman las constelaciones propiamente dichas) pero rotada en aproximadamente 180 grados, y muestra solo la zona sur de la eclíptica. Corresponde a latitudes negativas, por debajo de la línea recta horizontal de la parte superior, que representa a la eclíptica y señala la latitud eclíptica cero. En el panel de la derecha se muestra una ampliación de la misma imagen en la zona de mayor acercamiento entre el límite de la Ballena y la eclíptica.**

Debido al movimiento secular de precesión, con el transcurso de las épocas el Sol no coincide con las mismas constelaciones del cielo en los sucesivos solsticios y equinoccios, ni se mantiene en la misma posición con relación a las estrellas durante las cuatro estaciones. Así, una persona nacida el 23 de marzo unos 2.000 años atrás (aproximadamente cuando se sistematiza la astrología) lo hizo cuando el Sol estaba entre las estrellas de Aries (el Carnero). En la época actual, en esa fecha está en la constelación de Piscis.

La astronomía moderna, en síntesis, es capaz de explicar los movimientos de los cuerpos celestes que dieron pie a los relatos astrológicos, los cuales abarcan hoy un extenso abanico de géneros, desde la literatura y el mito tradicional hasta la superstición. El contenido y valor de esos relatos, sin embargo, está fuera del campo de competencia de la ciencia.

| Constelación | Fecha tradicional | Fecha actual | Número de días |
|---|---|---|---|
| Capricornio | 22 dic. – 21 ene. | 20 ene. – 15 feb. | 27 |
| Acuario | 22 ene. – 21 feb. | 16 feb. – 11 mar. | 24 |
| Piscis | 22 feb. – 21 mar. | 12 mar. – 18 abr. | 38 |
| Aries | 22 mar. – 21 abr. | 19 abr. – 13 may. | 25 |
| Tauro | 22 abr. – 21 may. | 14 may. – 21 jun. | 39 |
| Gemini | 22 may. – 21 jun. | 22 jun. – 20 jul. | 29 |
| Cáncer | 22 jun. – 21 jul. | 21 jul. – 10 ago. | 21 |
| Leo | 22 jul. – 21 ago. | 11 ago. – 16 sep. | 37 |
| Virgo | 22 ago. – 21 sep. | 17 sep. – 30 oct. | 44 |
| Libra | 22 sep. – 21 oct. | 31 oct. – 22 nov. | 23 |
| Escorpión | 22 oct. – 21 nov. | 23 nov. – 29 nov. | 7 |
| Ofiuco | | 30 nov. – 17 dic. | 18 |
| Sagitario | 22 nov. – 21 dic. | 18 dic. – 19 ene. | 33 |

**El horóscopo tradicional sugiere que el Sol emplea igual número de días (unos 30 días, más o menos) para recorrer cada una de las 12 constelaciones del zodíaco. Sin embargo, eso *no* es cierto. La mayor parte del tiempo, el Sol se ubica en la constelación de la Virgen o Virgo (unos 44 días en el año) y dedica tan sólo 7 días a recorrer la del Escorpión. Es por ello que los "verdaderos" escorpión deberían ser una minoría. Además, y como ya lo mencionamos en el texto, existe un período (entre el 30 de noviembre y el 17 de diciembre) en el que el Sol se ubica en la constelación de Ofiuco y –al menos en nuestro modesto conocimiento– jamás ninguno de nosotros se encontró con un *ofiuco* en toda su vida… (¡y ni hablar de un *cetus*! – ver el texto). La tabla indica las constelaciones zodiacales, junto a sus fechas tradicionales, los períodos del año en los que el Sol pasa por cada constelación y el número aproximado de días correspondiente.**

RECUADRO: *Una hipótesis "fantástica"*

Avatares políticos ocurridos durante el año 1302 condenaron a Dante, en ese entonces uno de los *priori* –o supremo magistrado– de Florencia, a vivir en el exilio por el resto de sus días. Dante inicia así un largo peregrinar por varias cortes del norte de Italia: Verona, Treviso, Padua… para terminar estableciéndose en la ciudad de Rávena bajo la protección de un poeta y amigo suyo que gobernaba la ciudad, Guido da Polenta. Será durante todos estos años de destierro que el poeta compondrá varias de sus obras más importantes. Entre éstas, encontramos *De vulgari eloquentia* (*Sobre la lengua vulgar*), *Convivio* (*El Banquete*), *De monarchia* (*Sobre la monarquía*), y *De situ et forma aque et terre* (*Sobre la ubicación y la forma de la esfera del agua y de la tierra*), esta última quizás no siempre considerada tan relevante entre los estudiosos de Dante, pero que aquí nos permitirá conocer algo más sobre los conocimientos "científicos" del poeta. Por último, Dante dará a luz su célebre *Commedia*, que llegaría a nuestros días como la *Divina Comedia*, y en cuyas páginas nos ofrece un testimonio elocuente de la dureza del conflicto político florentino que le tocó vivir.

Su obra *Sobre la ubicación y la forma de la esfera del agua y de la tierra* (1320), muchas veces más conocida bajo el nombre *Quaestio de aqua et terra* (*Cuestión del agua y de la tierra*), se trataría de la versión escrita de un discurso pronunciado por el poeta el 20 de enero de 1320 en el pequeño templo de Santa Elena, en Verona, frente al clero local. En éste, Dante explica los motivos por los cuales es posible que en ciertas partes del globo terrestre el elemento más pesado entre los cuatro elementos básicos –la tierra– logre sobresalir por encima del agua –más liviana– para formar los continentes, comprendidos éstos entre las "columnas de Hércules" –el estrecho de Gibraltar– y el río Ganges, que era la tierra firme donde Dios había asignado la morada de los hombres luego del pecado de Adán.

Recordemos que, de acuerdo a las enseñanzas de Aristóteles, los cuatro elementos básicos se organizaban en esferas concéntricas en el orden: tierra, agua, aire, fuego, desde el centro de la Tierra y hacia afuera, con la esfera del agua cubriendo totalmente al elemento tierra. Los motivos de la disertación de Dante no son del todo claros, pero podrían deberse a unas aclaraciones que se le pidieran al poeta sobre el último canto del Infierno. En ese cántico de la *Commedia*, Dante, bajo la excusa bíblica, explica la emersión terrestre como el resultado de la "caída de Lucifer" del cielo, que de esa manera habría desplazado grandes masas de tierra haciéndoles sobresalir por encima de la esfera del agua. En su disertación, Dante, enfrentado nuevamente con este problema cosmogónico, tácitamente parece negar sus escritos de la Divina Comedia y volcarse a una hipótesis diferente, aunque no menos "fantástica": el poeta esta vez explica la emersión de regiones de tierra sobre los océanos como el resultado de la *influencia* de las estrellas; es esta "influencia", vista quizás como una forma de "atracción", la que habría generado en el hemisferio boreal una suerte de "joroba" en la esfera de la tierra.

Mencionamos que Dante no creía en la influencia de los astros sobre la mente y el alma humanas; los astros, en cambio, sí podían alterar las cosas terrenas. Sería esta deformación "astral" de la esfera terrestre la que habría permitido el retroceso de las aguas y la existencia de la tierra firme, y, como consecuencia, también de la vida.

RECUADRO: *El movimiento de los cielos*

Todos sabemos que los días son más largos en verano y que en invierno oscurece más temprano. También recordamos con alegría el comienzo de la primavera y tenemos conciencia de que la caída de hojas amarillas indica el arribo del otoño. En astronomía, las cuatro estaciones (o mejor dicho, sus inicios) también tienen nombres apropiados: *solsticios* y *equinoccios*.

Al solsticio de diciembre se lo denomina *solsticio de verano austral* e indica que comienza el verano en el hemisferio sur. Es el día más largo del año y el día en el que el "arco aparente" que describe el Sol en el cielo, que con el pasar de los días se venía alargando, llega a su máxima longitud. A partir de esta fecha, dicho arco comenzará a decrecer cada vez más (y con ello también lo hará la "duración de los días") hasta llegar al *solsticio de invierno austral* (el 21 de junio aproximadamente) en el que tendremos el día más corto del año. En el hemisferio norte o boreal se invierten el sentido del crecimiento y las estaciones.

La palabra solsticio viene del Latín (*solstitium*) y significa "Sol quieto", por supuesto, *no* porque el astro deje de "surgir" y de "ocultarse" en el horizonte, sino porque las posiciones más altas en el cielo del mediodía, comparadas día tras día, llegados los solsticios, parecen permanecer idénticas: parece anularse la *variación* de las posiciones verticales máximas (en verano) y mínimas (en invierno) que alcanza el Sol en esos días.

En seis meses, entonces, se pasa de tener menos de 12 horas de luz solar (pleno invierno) a tener más de 12 horas en verano. Aquellos días del año en los que la cantidad de horas de luz y de sombra coinciden, cuando los días duran lo mismo que las noches para cualquier punto de la Tierra, se llaman equinoccios: *aequinoctium* en Latín, o sea "igual noche" para todo habitante de la Tierra. (Entre las constelaciones zodiacales, la que mejor refleja estos momentos del año es la constelación de *la balanza* –Libra– la cual, hace un par de miles de años simbolizaba el equilibrio entre el día y la noche en el equinoccio de primavera austral.) Tendremos entonces dos equinoccios: el de otoño austral (el 21 de marzo aproximadamente) cuando los días se acortan, y el de primavera austral (aproximadamente el 21 de septiembre) cuando los días se hacen cada vez más largos.

Todos los años, entonces, el solsticio de junio ocurre alrededor del día 21, y resulta ser que en ese momento el Sol se encuentra en el "borde" entre las constelaciones de Tauro y Gemini. Aquí –y más arriba– hemos escrito reiteradamente "alrededor" y "aproximadamente" porque no siempre solsticios y equinoccios caen exactamente los días 21. Para los habitantes del hemisferio sur, por ejemplo, el equinoccio de primavera y el solsticio de verano correspondientes al año 2007 fueron, respectivamente, el día 23 de septiembre a las 9h 51m TU y el día 22 de diciembre a las 6h 08m TU, ambos horarios expresados en Tiempo Universal (TU) – basta restar 3 horas para expresarlos en términos de la hora oficial argentina.
Sin embargo, miles de años atrás, el solsticio de invierno austral *no* coincidía con el Sol en la interface Tauro-Gemini (ver *tabla adjunta*), sino con el Sol entre las constelaciones de Gemini y Cáncer (el Cangrejo). Y esto es debido a un movimiento "muy especial" que posee el eje de rotación de la Tierra, y que llamamos *precesión*. Veamos ahora de qué se trata.

La precesión es un movimiento de rotación muy lento del eje de la Tierra en el espacio. Es que la Tierra en realidad no es una esfera perfecta, sino que, debido a su rotación, es más protuberante en el Ecuador que en los polos. Y la atracción gravitatoria, principalmente de la Luna, sobre este abultamiento ecuatorial genera una perturbación que hace que el eje de giro terrestre cambie su orientación en el espacio. A este cambio, a este movimiento de rotación que sufre el propio eje, se lo llama precesión y es similar (aunque muchísimo más lento) al que sufre el eje de un trompo en rotación cuando su propio peso lo comienza a desequilibrar para hacerlo caer: en efecto, al eje de la Tierra le lleva unos 26.000 años dar una vuelta completa (25.765 años para ser más precisos).
Como consecuencia de este cambio de orientación, los polos Norte y Sur celestes se "mueven" con respecto a las estrellas lejanas (en particular, la estrella Polaris *no* fue ni será siempre la estrella polar norte), y, en idéntica manera, las posiciones de solsticios y equinoccios también cambiarán, muy lentamente, su posición zodiacal. Si hoy el solsticio del verano boreal (e invierno austral) ocurre cuando el Sol está ubicado en el borde entre Gemini y Tauro, en el futuro –dentro de un par de decenas de siglos– ocurrirá cuando el Sol esté bien centrado sobre Tauro.

RECUADRO: *Sobre la astrología*

Es muy común −lamentablemente aún en nuestros días− crear confusión entre la astronomía y la astrología. Quizás, para muchos, esto se debe a no prestar demasiada atención al hablar, pero frecuentemente el motivo de la confusión es, simplemente, que la diferencia entre las dos "astro" no se conoce.

Podemos dar una definición básica de "la astronomía" como la ciencia que estudia los objetos celestes, como la Luna, el Sol, los planetas, las estrellas, las galaxias y demás megaestructuras del cielo (aunque sin dudas podríamos hacer esta definición algo más completa). Es una ciencia en el mismo sentido que lo son la física, la química o la biología: procura entender la índole y el funcionamiento del mundo material, específicamente de los cuerpos celeste. Si bien tiene una larga historia y fue practicada en distintas formas por casi todos los pueblos, hoy es una ciencia que realiza observaciones y mediciones, formula hipótesis y construye teorías, en una palabra, que crea conocimiento en el marco de la tradición científica moderna de Occidente.

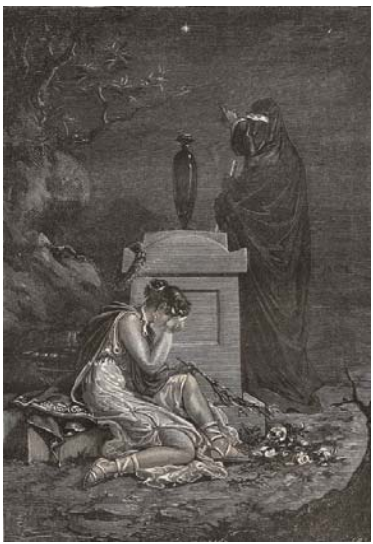

**Dentro de las ideas astrológicas de la antigüedad, Saturno –que brillaba en el cielo nocturno con su tinte plomizo y poseía el movimiento más lento de entre todos los planetas conocidos de la época– tenía una influencia nefasta y estaba asociado a los mayores dolores de la humanidad. Ya en el siglo I a.C., el poeta latino Albio Tibulo, en la tercera elegía de su primer libro de poemas (versos 17-18) rememoraba cómo cuando, temeroso de partir a la guerra, él "veía malos augurios en el vuelo de las aves, o invocaba que el día de Saturno [*Saturni*] debía permanecer en casa". Y es esta, aparentemente, la primera referencia escrita al sábado ("día de Saturno") como día de mal agüero, y poco conveniente para emprender un viaje. Imagen de *Astronomie Populaire*, 1880, de Camille Flammarion.**

Por otro lado, definir "la astrología" no es tan fácil. Está constituida por mitos y tradiciones, por relatos que exploran la condición humana en cuya base se encuentra la creencia, muy difundida en la actualidad y en todas las épocas, de que las vicisitudes humanas (y quizás las de todo organismo viviente), incluidas las personalidades de mujeres y hombres están influidas o determinadas por las posiciones, con relación a las estrellas, en que se veían los planetas, la Luna y el Sol en determinados momentos, particularmente en el del nacimiento de cada uno. Tales posiciones varían para los habitantes de la Tierra debido a las órbitas que este y los restantes planetas describen alrededor del Sol, pero se mantienen dentro de una franja del cielo llamada el Zodíaco, dividida en dice porciones, cada una de las cuales determina uno de los llamados *signos del Zodíaco*.

Parientes naturales de la astrología son la astrolatría (la adoración de los astros) y la astromancia (la adivinación por los astros). La primera no tiene mayormente lugar en el mundo de hoy, excepto quizás en épocas estivales, cuando vemos miles de seres "adorando" al Sol para tomar un poco de color. Por el contrario, la astromancia es una de las cosas que más entretiene a los lectores de periódicos y revistas (el horóscopo).

Más allá de las frases que escribimos arriba, en estas dos "definiciones" hay dos palabras importantes que vale la pena resaltar: la primera es "ciencia", la otra "creencia". Una ciencia evoluciona, se equivoca y se corrige, y se halla en continuo proceso de perfeccionamiento. Una creencia, en cambio, se mantiene intacta, y en el caso de la astrología, idéntica a como era hace varios miles de años cuando fue inventada. Hoy sabemos bien que la astrología es no-científica en su funcionamiento; los astrólogos jamás se equivocan o, en otros términos, siempre tienen razón *a posteriori*. Ahora bien, lo que define una ciencia es su capacidad de estar equivocada, a fin de progresar. El lenguaje astrológico no se comprende en las ciencias; la racionalidad científica no tiene lugar en la astrología. ¿Cómo se pueden refutar racionalmente aquellos razonamientos que no lo son?

## Agradecimientos

## Lecturas sugeridas